\documentstyle[12pt,subeqnarray]{article}
\begin{document}

\date{July 27, 1993}

\title{\vspace*{-1.5cm} A General Limitation\break
       on Monte Carlo Algorithms\break
       of Metropolis Type}
\author{
  \\
  {\small Sergio Caracciolo}              \\[-0.2cm]
  {\small\it Scuola Normale Superiore and INFN -- Sezione di Pisa}  \\[-0.2cm]
  {\small\it Piazza dei Cavalieri}        \\[-0.2cm]
  {\small\it Pisa 56100, ITALIA}          \\[-0.2cm]
  {\small Internet: {\tt CARACCIO@UX1SNS.SNS.IT}}     \\[-0.2cm]
  {\small Bitnet:   {\tt CARACCIO@IPISNSVA.BITNET}}   \\[-0.2cm]
  {\small Hepnet/Decnet:   {\tt 39198::CARACCIOLO}}   \\[-0.2cm]
  \\[-0.1cm]  \and
  {\small Andrea Pelissetto\thanks{
      Address until August 31, 1994.
      Permanent address:
      Dipartimento di Fisica and INFN -- Sezione di Pisa,
      Universit\`a degli Studi di Pisa,
      Pisa 56100, ITALIA.
      Internet: {\tt PELISSET@SUNTHPI1.DIFI.UNIPI.IT};
      Bitnet:   {\tt PELISSET@IPISNSVA.BITNET};
      Hepnet/Decnet:   {\tt 39198::PELISSETTO}.   }  }    \\[-0.2cm]
  {\small Alan D. Sokal}                  \\[-0.2cm]
  {\small\it Department of Physics}       \\[-0.2cm]
  {\small\it New York University}         \\[-0.2cm]
  {\small\it 4 Washington Place}          \\[-0.2cm]
  {\small\it New York, NY 10003 USA}      \\[-0.2cm]
  {\small Internet: {\tt PELISSET@MAFALDA.PHYSICS.NYU.EDU},
                    {\tt SOKAL@ACF4.NYU.EDU} }        \\[-0.2cm]
  {\protect\makebox[5in]{\quad}}  % To force authors' names to be written
                                  %   vertically, one above another.
                                  % (\author seems to put them side-by-side
                                  %   if there is room.)
  \\
}
\vspace{0.5cm}

\maketitle
\thispagestyle{empty}   % Suppress page number on front page.

\vspace{0.2cm}

\begin{abstract}
We prove that for any Monte Carlo algorithm of Metropolis type,
the autocorrelation time of a suitable ``energy''-like
observable is bounded below by a multiple of the corresponding
``specific heat''.
This bound does not depend on whether the proposed moves are
local or non-local;
it depends only on the distance between the desired probability distribution
$\pi$ and the probability distribution $\pi^{(0)}$ for which the proposal
matrix satisfies detailed balance.
We show, with several examples,
that this result is particularly powerful when applied
to non-local algorithms.
\end{abstract}

\vspace{0.5 cm}
\noindent
{\bf PACS number(s):}  02.70.Lq, 02.50.Ga, 05.50.+q, 11.15.Ha

\clearpage

\newcommand{\be}{\begin{equation}}
\newcommand{\ee}{\end{equation}}
\newcommand{\<}{\langle}
\renewcommand{\>}{\rangle}
\newcommand{\para}{\|}
\renewcommand{\perp}{\bot}

%\ltapprox and \gtapprox produce > and < signs with twiddle underneath
\def\spose#1{\hbox to 0pt{#1\hss}}
\def\ltapprox{\mathrel{\spose{\lower 3pt\hbox{$\mathchar"218$}}
 \raise 2.0pt\hbox{$\mathchar"13C$}}}
\def\gtapprox{\mathrel{\spose{\lower 3pt\hbox{$\mathchar"218$}}
 \raise 2.0pt\hbox{$\mathchar"13E$}}}
\def\inapprox{\mathrel{\spose{\lower 3pt\hbox{$\mathchar"218$}}
 \raise 2.0pt\hbox{$\mathchar"232$}}}

\def\half{ {{1 \over 2 }}}
\def\scra{{\cal A}}
\def\scrc{{\cal C}}
\def\scre{{\cal E}}
\def\scrf{{\cal F}}
\def\scrm{{\cal M}}
\newcommand{\scrmvec}{\vec{\cal M}}
\def\scrp{{\cal P}}
\def\scrs{{\cal S}}
\def\scrt{{\cal T}}
\def\ttens{{\stackrel{\leftrightarrow}{T}}}
\def\scrttens{{\stackrel{\leftrightarrow}{\cal T}}}
\def\scrv{{\cal V}}
\def\scrw{{\cal W}}
\def\scry{{\cal Y}}
\def\tauss{\tau_{int,\,\scrm^2}}
\def\taux{\tau_{int,\,{\cal M}^2}}
\newcommand{\taum}{\tau_{int,\,\vec{\cal M}}}
\def\taue{\tau_{int,\,{\cal E}}}
\newcommand{\imag}{\mathop{\rm Im}\nolimits}
\newcommand{\real}{\mathop{\rm Re}\nolimits}
\newcommand{\tr}{\mathop{\rm tr}\nolimits}
\newcommand{\sgn}{\mathop{\rm sgn}\nolimits}
\newcommand{\codim}{\mathop{\rm codim}\nolimits}
\def\textprime{{${}^\prime$}}
\newcommand{\longto}{\longrightarrow}
\def\var{ \hbox{var} }
\newcommand{\gtilde}{ {\widetilde{G}} }
\newcommand{\USp}{ \hbox{\it USp} }
\newcommand{\CP}{ \hbox{\it CP\/} }
\newcommand{\QP}{ \hbox{\it QP\/} }
\def\hboxscript#1{ {\hbox{\scriptsize\em #1}} }

\newcommand{\plotdot}{\makebox(0,0){$\bullet$}}
\newcommand{\plotsmalldot}{\makebox(0,0){{\footnotesize $\bullet$}}}

\def\bsigma{\mbox{\protect\boldmath $\sigma$}}
\def\btau{\mbox{\protect\boldmath $\tau$}}
  % \boldmath is fragile, and without the \protect we get screwed when
  % we try to use \bsigma in a \caption.
\def\br{{\bf r}}

\newcommand{\reff}[1]{(\ref{#1})}

%\font\specialroman=msym10 scaled\magstep1  % 12-point Special Roman (caps
%%only)
%\font\sevenspecialroman=msym7              % 7-point Special Roman (caps only)
%\def\zed{\hbox{\specialroman Z}}
%\def\szed{\hbox{\sevenspecialroman Z}}
%\def\R{\hbox{\specialroman R}}
%\def\sR{\hbox{\sevenspecialroman R}}
%\def\N{\hbox{\specialroman N}}
%\def\C{\hbox{\specialroman C}}
%\def\Q{\hbox{\specialroman Q}}
%\renewcommand{\emptyset}{\hbox{\specialroman ?}}
\newcommand{\zed}{{\bf \rm Z}}
\newcommand{\R}{\hbox{{\rm I}\kern-.2em\hbox{\rm R}}}
\font\srm=cmr7 		% to get seven roman
\def\szed{\hbox{\srm Z\kern-.45em\hbox{\srm Z}}}
\def\sR{\hbox{{\srm I}\kern-.2em\hbox{\srm R}}}
\def\C{{\bf C}}

% \font\german=eufm10 scaled\magstep1	% 12-point Euler Fraktur (German)
% \def\germang{\hbox{\german g}}
% \def\germansu{\hbox{\german su}}

% \font\amssymbol=msxm10 scaled \magstep1  % Another AMS symbol font
% \def\transversal{\hbox{\amssymbol t}}  % THERE MAY BE A BETTER SYMBOL.

\newtheorem{theorem}{Theorem}[section]
\newtheorem{corollary}[theorem]{Corollary}
\newtheorem{lemma}[theorem]{Lemma}
\def\proof{\bigskip\par\noindent{\sc Proof.\ }}
\def\qed{\hbox{\hskip 6pt\vrule width6pt height7pt depth1pt \hskip1pt}\bigskip}

%
% Array for subscripts
%
\newenvironment{sarray}{
          \textfont0=\scriptfont0
          \scriptfont0=\scriptscriptfont0
          \textfont1=\scriptfont1
          \scriptfont1=\scriptscriptfont1
          \textfont2=\scriptfont2
          \scriptfont2=\scriptscriptfont2
          \textfont3=\scriptfont3
          \scriptfont3=\scriptscriptfont3
        \renewcommand{\arraystretch}{0.7}
        \begin{array}{l}}{\end{array}}

\newenvironment{scarray}{
          \textfont0=\scriptfont0
          \scriptfont0=\scriptscriptfont0
          \textfont1=\scriptfont1
          \scriptfont1=\scriptscriptfont1
          \textfont2=\scriptfont2
          \scriptfont2=\scriptscriptfont2
          \textfont3=\scriptfont3
          \scriptfont3=\scriptscriptfont3
        \renewcommand{\arraystretch}{0.7}
        \begin{array}{c}}{\end{array}}

Forty years ago, Metropolis {\em et al.}\/ \cite{Metropolis}
introduced a general method for constructing dynamic Monte Carlo algorithms
(= Markov chains \cite{Markov_books})
that satisfy detailed balance for a specified probability distribution $\pi$.
In this note we would like to point out a general limitation on {\em all}\/
algorithms of Metropolis type.
We prove that the autocorrelation time of a suitable ``energy''-like
observable is bounded below by a multiple of the corresponding
``specific heat''.
This bound does not depend on whether the proposed moves are
local or non-local;
it depends only on the distance between the desired probability distribution
$\pi$ and the probability distribution $\pi^{(0)}$ for which the proposal
matrix satisfies detailed balance.

Let us begin by recalling
the general Metropolis {\em et al.}\/ \cite{Metropolis} method,
as slightly generalized by Hastings \cite{Hastings}.
We use the notation of a discrete (finite or countably infinite)
state space $S$, but the same considerations apply with minor modifications
to a general measurable state space.
%% {\bf CHANGE BELOW TO USE ARROW NOTATION FOR TRANSITION MATRICES AND
%%     ACCEPTANCE FRACTION?????}
Let $P^{(0)} = \{ p^{(0)}_{xy} \}$ be an arbitrary %%%irreducible
transition matrix on $S$.  We call $P^{(0)}$ the
{\em proposal matrix}\/, and use it to generate proposed
moves $x \to y$ that will then be accepted or rejected
with probabilities $a_{xy}$ and $1-a_{xy}$, respectively.
If a proposed move is rejected, we make a ``null transition'' $x \to x$.
The transition matrix $P = \{ p_{xy} \}$ of the full algorithm is thus
\be
   p_{xy}   \;=\;
   \cases{ p^{(0)}_{xy} \, a_{xy}  & for $x \neq y$   \cr
           \noalign{\vskip 2pt}
           p^{(0)}_{xx} \,+\, \sum\limits_{z\neq x} p^{(0)}_{xz} \, (1-a_{xz})
                                   & for $x=y$   \cr
         }
\ee
where of course we must have $0 \le a_{xy} \le 1$ for all $x,y$.
It is easy to see that $P$ satisfies detailed balance for $\pi$
if and only if
\be
   {a_{xy} \over a_{yx}}   \;=\;
   { \pi_y \, p^{(0)}_{yx}   \over   \pi_x \, p^{(0)}_{xy} }
\ee
for all pairs $x \neq y$.
But this is easily arranged:  just set
\be
   a_{xy}   \;=\;  F\!\left(
                      { \pi_y \, p^{(0)}_{yx}   \over   \pi_x \, p^{(0)}_{xy} }
                      \right)   \;,
  \label{acceptance}
\ee
where $F\colon\; [0,+\infty] \to [0,1]$ is any function satisfying
\be
  {F(z) \over F(1/z)}  \;=\;  z  \qquad\hbox{for all } z .
 \label{a_condition}
\ee
The choice suggested by Metropolis {\em et al.}\/ \cite{Metropolis} is
\be
   F_{\hboxscript{Metr}}(z) \;=\; \min(z,1)   \;.
\ee
Other choices of $F$ are possible, but it is easy to see that they all
must satisfy the {\em inequality}\/
\be
   F(z) \;\le\; \min(z,1)   \;.
 \label{upper_bound_a}
\ee
Of course, it is still necessary to check that $P$ is irreducible (= ergodic);
this is usually straightforward.

Note that if the proposal matrix $P^{(0)}$ happens to {\em already}\/
satisfy detailed balance for $\pi$, then we have
$\pi_y  p^{(0)}_{yx}  /  \pi_x  p^{(0)}_{xy} = 1$,
so that $a_{xy} =1$ (if we use the Metropolis choice of $F$)
and $P = P^{(0)}$.  On the other hand, no matter what $P^{(0)}$ is,
we obtain a matrix $P$ that satisfies detailed balance for $\pi$.
So the Metropolis procedure can be thought of as a
prescription for minimally modifying a given transition matrix $P^{(0)}$
so that it satisfies detailed balance for $\pi$.

Let us now assume that $P^{(0)}$ satisfies detailed balance for some
probability measure $\pi^{(0)}$;  in practice this is virtually always
the case.
We then define an energy-like observable $H$ by
\be
   H(x)   \;=\;
   \cases{ -\log(\pi_x/\pi^{(0)}_x)    &  if $\pi_x > 0$  \cr
           +\infty                     &  if $\pi_x = 0$  \cr
         }
\ee
%% {\bf DISCUSS EARLIER VARIOUS POSSIBILITIES OF ZERO PROBABILITIES IN
%%      $\pi$ AND/OR $\pi^{(0)}$;  EXCLUDE SOME WITHOUT LOSS OF GENERALITY????}
The point is that $H$ is the ``energy'' of the probability distribution $\pi$
{\em relative to}\/ $\pi^{(0)}$.

The heart of our argument is the following upper bound on the
mean-square change in energy in a single step of the Metropolis algorithm:

\bigskip
\par\noindent
{\bf Proposition.}  In the situation described above, we always have
\be
   \< (\Delta H)^2 \>  \;\equiv\;
   \sum\limits_{x,x'}  \pi_x \, p_{xx'} \, [H(x') - H(x)]^2
   \;\le\;
   {8 \over e^2} f_+
   \;\le\;
   {8 \over e^2}
   \;,
 \label{ms_deltaE}
\ee
where
\be
   f_+   \;\equiv\;
 \!\sum\limits_{\begin{scarray}
                  x,x' \\  H(x') > H(x)
                \end{scarray}}
 \!\pi_x \, p^{(0)}_{xx'}
   \;\le\;  1
\ee
is the fraction (in equilibrium) of proposals that would strictly increase
the energy.

\proof
Since $P$ satisfies detailed balance for $\pi$,
the summand in \reff{ms_deltaE} is symmetric under $x \leftrightarrow x'$.
Therefore it suffices to consider the terms for which $H(x') > H(x)$,
and to multiply the result by 2.
(The terms having $H(x') = H(x)$ of course make no contribution to the sum.)

If $H(x') > H(x)$, we have
$a_{xx'} \le e^{-[H(x')-H(x)]}$ by \reff{acceptance} and
\reff{upper_bound_a}.
Therefore
\begin{eqnarray}
   \sum\limits_{\begin{scarray}
                  x,x' \\  H(x') > H(x)
                \end{scarray}}
   \pi_x \, p_{xx'} \, [H(x') - H(x)]^2
  & = &
   \sum\limits_{\begin{scarray}
                  x,x' \\  H(x') > H(x)
                \end{scarray}}
   \pi_x \, p^{(0)}_{xx'} \, a_{xx'} \, [H(x') - H(x)]^2   \nonumber \\[2mm]
  & \le &
   \sum\limits_{\begin{scarray}
                  x,x' \\  H(x') > H(x)
                \end{scarray}}
   \pi_x \, p^{(0)}_{xx'} \, e^{-[H(x')-H(x)]} \, [H(x') - H(x)]^2
                                                           \nonumber \\[2mm]
  & \le &
   {4 \over e^2}  f_+
\end{eqnarray}
since $z^2 e^{-z} \le 4/e^2$ for all $z \ge 0$.
\qed

The physical intuition behind this proof is simple:
Proposed moves having a large energy change $\Delta H > 0$
have an exponentially small acceptance probability,
so the mean-square energy increase $\< (\Delta H)_+^2 \>$
in a single Metropolis step is at most of order 1.
Proposed moves having a energy change $\Delta H < 0$
are connected to those with $\Delta H > 0$ by detailed balance:
when proposed they are accepted, but if $|\Delta H|$ is large
they are only rarely proposed.
The result is that the mean-square energy change in either direction
is at most of order 1.

Let us now recall the definitions of autocorrelation functions
and autocorrelation times \cite{Sokal_Lausanne}:
If $A$ is a real-valued function defined on the state space $S$
(i.e.\ a real-valued observable),
we define its unnormalized autocorrelation function (in equilibrium) by
\begin{subeqnarray}
C_{AA} (t)   & \equiv &   \< A_s  A_{s+t} \> - \mu_A^2       \\[2mm]
             &    =   &   \sum\limits_{x,y}
                  A(x) \, [ \pi_x (P^{|t|})_{xy} - \pi_x \pi_y ] \, A(y)   \;.
\end{subeqnarray}
The corresponding normalized autocorrelation function is
\be
  \rho_{AA} (t) \;\equiv\; C_{AA} (t) / C_{AA} (0) \;.
\ee
The integrated and exponential autocorrelation times are then defined by
\begin{eqnarray}
 \tau_{int,A}   & = &   \half \sum_{t=-\infty}^{\infty}  \rho_{AA} (t)  \\[2mm]
 \tau_{exp,A}   & = &   \limsup\limits_{t \to \infty} \,
                                {|t| \over -  \log  | \rho_{AA} (t)|}   \\[2mm]
 \tau_{exp}     & = &  \sup\limits_A  \, \tau_{exp,A}
\end{eqnarray}
Some simple identities are worth noting:
\begin{subeqnarray}
 \label{identities}
   C_{AA}(0)   & = &   \< A^2 \> _\pi   \,-\,   \< A \>^2 _\pi        \\[2mm]
   C_{AA}(1)   & = &   C_{AA}(0)  \,-\,
     {1 \over 2} \sum\limits_{x,x'}  \pi_x \, p_{xx'} \, [A(x') - A(x)]^2
\end{subeqnarray}
Also, from detailed balance combined with the spectral theorem one can
deduce the following inequalities:
\begin{eqnarray}
   \tau_{int,A}   & \ge &
      {1 \over 2} \,   { 1 + \rho_{AA}(1)   \over   1 - \rho_{AA}(1) }
                                                     \label{ineq1}    \\[2mm]
   \tau_{exp}   \;\ge\;  \tau_{exp,A}   & \ge &   -1 / \log |\rho_{AA}(1)|
                                                     \label{ineq2}
\end{eqnarray}
(see e.g.\ \cite[Appendix A]{CPS_90}).

With these preliminaries, the following theorem is an immediate consequence
of the Proposition:

\bigskip
\par\noindent
{\bf Theorem.}  Under the preceding hypotheses, we have
\begin{subeqnarray}
    \tau_{int,H}   & \ge &  {e^2 \over 4} {\var(H) \over f_+}
                                                       \,-\,  \half  \\[2mm]
    \tau_{exp}     & \ge &  -1 / \log( 1 - 4f_+/e^2 \var(H))
\end{subeqnarray}
where $\var(H) \equiv \< H^2 \> _\pi - \< H \>_\pi ^2$.

\proof
{}From the Proposition together with \reff{identities}, we get
\be
   \rho_{HH}(1)   \;\equiv\;
   { C_{HH}(0) \over C_{HH}(1) }
   \;\ge\;   1  \,-\,  {4 \over e^2 \var(H)}
\ee
Now use \reff{ineq1} and \reff{ineq2}.
\qed

Again the physical intuition is simple:
The mean-square energy change per Me\-trop\-o\-lis step is at most of order 1.
On the other hand, in order to sample adequately the probability distribution
$\pi$, the Markov chain must traverse an energy distribution
of width $\sim \var(H)^{1/2}$.
This takes a time of order $(\var(H)^{1/2})^2 \sim \var(H)$.

\bigskip

{\bf Example 1.}
{\em Single-site Metropolis algorithm.}\/
Here $\pi^{(0)}$ is the {\em a priori}\/ measure for the spins,
and $H$ is the full Hamiltonian.
$P^{(0)}$ selects a spin at random and proposes to update it
in some way that satisfies detailed balance for $\pi^{(0)}$.
We have $\var(H) = V C_h$,
where $V$ is the volume and $C_h$ is the specific heat.
So the Theorem shows that
\be
   \tau_{int,H}, \tau_{exp,H}   \;\gtapprox\;
   V C_h   \;,
\ee
where time is here measured in hits of a single site;
or equivalently $\tau \gtapprox C_h$ when time is measured in ``sweeps''.
This is a well-known result.
However, it is a rather poor bound because the energy,
being a {\em short}\/-distance observable,
has a rather weak overlap with the slowest ({\em long}\/-wavelength) modes
of this local dynamics.
(A much stronger bound can be obtained by using the magnetization $\scrm$
 rather than the energy as the trial function:
 one gets $\tau_{int,\scrm}, \tau_{exp,\scrm} \gtapprox V\chi$,
 where $\chi$ is the susceptibility \cite{chi_bound,Sokal_Lausanne}.)

\bigskip

The real power of the Theorem comes when it is applied to
{\em non-local}\/ algorithms:
it still yields $\tau \gtapprox V C_h$, but now the unit of time
(a ``hit'' of $P^{(0)}$) is a non-local move which costs a CPU time $\gg 1$.
As a result, several algorithms which {\em a priori}\/ look promising
must in fact perform rather poorly:

\bigskip

{\bf Example 2.}
{\em $q$-state Potts model with mixed ferromagnetic/antiferromagnetic
     interaction}\/  \cite{Potts_mixed_refs}.
The purely ferromagnetic Potts model can be simulated very efficiently
by the Swendsen-Wang (SW) algorithm \cite{Swendsen_87,Li_89}
or its single-cluster (1CSW) variant
\cite{Wolff_89a,1CSW_papers},
but these algorithms do not extend easily to the
mixed ferromagnetic/antiferromagnetic case.
One might therefore try using the SW or 1CSW algorithm for the
ferromagnetic part of the Hamiltonian
as a Metropolis proposal for the full theory.
Thus, let $\pi^{(0)}$ (resp.\ $\pi$) be the Gibbs measure for the
ferromagnetic (resp.\ full) theory, so that $H$ is the antiferromagnetic
part of the Hamiltonian.
Let $P^{(0)}$ be {\em any}\/ algorithm that satisfies detailed balance
for $\pi^{(0)}$ (for example, SW or 1CSW);
and let $P$ be the corresponding Metropolis algorithm for $\pi$.
One expects $\var(H)$ to behave near criticality as
$\sim J_{\hboxscript{af}}^2 V C_h$,
where $J_{\hboxscript{af}}$ is the antiferromagnetic coupling.
So the Theorem shows that
\be
   \tau_{int,H}, \tau_{exp,H}   \;\gtapprox\;
   J_{\hboxscript{af}}^2 V C_h   \;,
\ee
where time is here measured in hits of $P^{(0)}$.
For SW (resp.\ 1CSW), each hit takes a CPU time of order $V$
(resp.\ $\chi$).
So the proposed algorithm must perform quite poorly,
except when $J_{\hboxscript{af}}$ is very small \cite{z_for_example_2}.

\bigskip

{\bf Example 3.}
{\em $d=3$ Heisenberg model with topological term}\/ \cite{Lau_89}.
The ferromagnetic Heisenberg model can be simulated very efficiently
by the Wolff embedding algorithm
\cite{Wolff_89a,Wolff_embedding_papers}
using either SW or 1CSW moves to update the induced Ising model
\cite{Holm_93a+b}.
The topological term seems difficult to incorporate into the
cluster-algorithm framework,
but one might try using the SW or 1CSW algorithm for the
ferromagnetic two-body part of the Hamiltonian
as a Metropolis proposal for the full theory.
(The intuitive idea is that a 1CSW move is likely to make a modest
 change in the topological-charge field, so the acceptance rate
 should be reasonable.)
Thus, let $\pi^{(0)}$ (resp.\ $\pi$) be the Gibbs measure for the
ferromagnetic (resp.\ full) theory, so that $H$ is the topological term.
Let $P^{(0)}$ be {\em any}\/ algorithm that satisfies detailed balance
for $\pi^{(0)}$ (for example, SW or 1CSW);
and let $P$ be the corresponding Metropolis algorithm for $\pi$.
One expects
$\var(H)$ to behave near criticality as $\sim J_{top}^2 V C_h$,
where $J_{top}$ is the topological coupling \cite{Holm_93c};
and it is known that $C_h \to {\rm const} > 0$ at criticality
(since $\alpha < 0$).
So the Theorem shows that
\be
   \tau_{int,H}, \tau_{exp,H}   \;\gtapprox\;
   J_{top}^2 V \;,
\ee
where time is here measured in hits of $P^{(0)}$.
For SW (resp.\ 1CSW), each hit takes a CPU time of order $V$
(resp.\ $\chi$).
So the proposed algorithm must perform quite poorly,
except when $J_{top}$ is very small.

\bigskip

{\bf Example 4.}
{\em Self-avoiding walk with nearest-neighbor interaction.}\/
Fix an integer $N$, and
let $S$ be the space of all $N$-step self-avoiding walks on some specified
lattice.  Let $\pi^{(0)}$ be the probability measure that gives equal weight
to each element of $S$.  Then define the probability measure $\pi$ by
\be
   \pi_\omega   \;=\;
      Z(\epsilon)^{-1} \, e^{-\epsilon M(\omega)} \, \pi^{(0)}_\omega
   \;,
 \label{sawnn}
\ee
where $M(\omega)$ is the number of non-bonded nearest-neighbor contacts
in the walk $\omega$.
Let $P^{(0)}$ be {\em any}\/ algorithm that satisfies detailed balance
for $\pi^{(0)}$ (e.g.\ the pivot algorithm \cite{Lal_69,Madras_88});
and let $P$ be the corresponding Metropolis algorithm for \reff{sawnn}.
Then the Theorem shows that
\be
   \tau_{int,M}, \tau_{exp,M}   \;\gtapprox\;
   \epsilon^2 \var_\pi(M) / f   \;,
  \label{sawnn_bound}
\ee
where $f$ is the fraction of proposals $p^{(0)}_{\omega \omega'}$
with $\omega' \neq \omega$
(e.g.\ the fraction of proposed pivot moves that preserve self-avoidance).
And we expect $\var_\pi(M) \approx N C(\epsilon)$,
where the ``specific heat per step'' $C(\epsilon)$
is everywhere nonzero and
diverges like $(\epsilon - \epsilon_\theta)^{-\alpha_\theta}$
at the theta (tricritical) point.

For the pivot algorithm, the bound \reff{sawnn_bound} is a rather weak
result:  in fact we expect that $\tau_{int,M}, \tau_{exp,M} \sim N/f$
even for $\epsilon = 0$, because $M$ is a ``primarily local'' observable
\cite{Madras_88}.
But \reff{sawnn_bound} does show that for $\epsilon \neq 0$
(and in particular for $\epsilon \to \epsilon_\theta$)
the difficulties cannot be avoided by using a different proposal $P^{(0)}$;
they are inherent in the Metropolis method
with this choice of $\pi^{(0)}$ \cite{footnote_re_conjecture}.

\bigskip

We conclude by noting that the Metropolis {\em et al.}\/ method
is often applied indirectly:
we define transition matrices $P_1,\ldots,P_n$ by the Metropolis method,
and we then execute either
$P = \sum_{i=1}^n \lambda_i P_i$ for some weights $\lambda_i \ge 0$
(``random updating'')
or else $P = P_1 \cdots P_n$ (``sequential updating'').
The first case can easily be handled by our method.
The second case is more subtle, because typically $P$ does not
satisfy detailed balance \cite{remark_re_detailed_balance};
but the bound is almost certainly correct in order of magnitude,
except in special situations like ``successive overrelaxation''
\cite{Adler_LAT88}.

\bigskip

The authors wish to thank Giovanni Ferraro for a helpful discussion.
One of us (A.D.S.) wishes to thank the Scuola Normale Superiore
for hospitality while this work was being carried out.
The authors' research was supported in part by
the Consiglio Nazionale delle Ricerche (S.C.\ and A.P.),
the Istituto Nazionale di Fisica Nucleare (S.C., A.P.\ and A.D.S.),
U.S.\ Department of Energy contract DE-FG02-90ER40581 (A.D.S.),
U.S.\ National Science Foundation grants DMS-8911273 and DMS-9200719 (A.D.S.),
and NATO Collaborative Research Grant CRG 910251 (S.C.\ and A.D.S.).

%
%
%%%%%%%%%%%%   references  %%%%%%%%%%%%%%%%%%%%%%%%
%

\clearpage


\begin{thebibliography}{99}
\bibitem{z_for_example_2}   In conventional CPU units of ``hits per site''
(equivalent to ``sweeps'' of a local algorithm), we have
$\tau \gtapprox V C_h$ (resp.\ $\chi C_h$).
In terms of dynamic critical exponents, this means
$z_{CPU} \ge  d + (\max(\alpha,0)/\nu)$
[resp.\ $(\gamma + \max(\alpha,0))/\nu$].
\bibitem{remark_re_detailed_balance}
Of course, $P = P_1 \cdots P_n$ does leave $\pi$ invariant,
which is all that matters.
\end{thebibliography}
\end{document}